\documentclass[11pt,a4paper,oneside]{amsart}
\usepackage[a4paper,margin=2.5cm]{geometry}
\usepackage[dutch,english]{babel}
\usepackage{graphicx,caption,subcaption}                         
\usepackage[pdftex,hyperfootnotes=false,colorlinks=true,linkcolor=blue,%
citecolor=purple,filecolor=magenta,urlcolor=cyan]{hyperref}
\usepackage{amsmath,amssymb,amscd,amsthm,amsfonts,anyfontsize,dsfont,%
enumerate,fix-cm,layout,lipsum,lpic,mathrsfs,mdwlist,pigpen,stmaryrd,%
tensor,thmtools,tikz,txfonts,xspace}
\usepackage[all]{xy}
\usepackage[oldstyle]{libertine}%
\usepackage[T1]{fontenc}
\usepackage[utf8]{inputenc}
\usepackage{csquotes}
\makeatletter
\renewcommand*\libertine@figurestyle{LF}
\makeatother
\usepackage[libertine,libaltvw,liby]{newtxmath}
\makeatletter
\renewcommand*\libertine@figurestyle{OsF}
\makeatother
\usepackage{mathtools}
\usepackage[noabbrev]{cleveref}
\expandafter\def\csname ver@etex.sty\endcsname{3000/12/31}

\usepackage{autonum}

\usepackage{hyperref}

\theoremstyle{plain}                          
\newtheorem{theorem}{Theorem}[subsection]
    
\newtheorem{lemma}[theorem]{Lemma}
\newtheorem{corollary}[theorem]{Corollary}
 
\theoremstyle{definition}
\newtheorem{definition}[theorem]{Definition}
\newtheorem{prop-defin}[theorem]{Proposition-definition} 

\theoremstyle{remark}

\begin{document}
\title{Ground states of Nicolai and $\mathbb{Z}_2$ Nicolai models}

\author{Ruben La}
\address[Ruben La]{Korteweg de Vries Institute for Mathematics and Institute of Physics, University of Amsterdam}
\email{ruben.la@outlook.com}

\author{Kareljan Schoutens}
\address[Kareljan Schoutens]{Institute of Physics and QuSoft, University of Amsterdam}
\email{c.j.m.schoutens@uva.nl}

\author{Sergey Shadrin}
\address[Sergey Shadrin]{Korteweg de Vries Institute for Mathematics, University of Amsterdam}
\email{s.shadrin@uva.nl}

\begin{abstract} 
We derive explicit recursions for the ground state generating functions of the one-dimensional Nicolai model and $\mathbb{Z}_2$ Nicolai model. Both are examples of lattice models with $\mathcal{N}=2$ supersymmetry. The relations that we obtain for the $\mathbb{Z}_2$ model were numerically predicted by Sannomiya, Katsura, and Nakayama.
\end{abstract}

\maketitle

\tableofcontents

 

\section{Introduction}

More than forty years ago, Hermann Nicolai proposed a lattice model featuring $\mathcal{N}=2$ supersymmetry \cite{Nicolai}. It was an early example of a realization of $\mathcal{N}=2$ supersymmetric quantum mechanics with an underlying spatial lattice structure. Supersymmetric quantum mechanics arises as soon as a quantum mechanical hamiltonian $H$ can be written as
\[
H = \{ Q, Q^\dagger \}
\]
with supercharges $Q$ and $Q^\dagger$ 
satisfying $Q^2 = (Q^\dagger)^2 = 0$. An easy consequence is that both supercharges commute with $H$,
\[
[ Q, H ]  = 0, \qquad [ Q^\dagger, H] = 0 .
\]
This algebraic structure implies that eigenstates of the hamiltonian, satisfying $H|\psi\rangle=E|\psi\rangle$, can be organized into doublets $\{|\psi\rangle,Q^\dagger |\psi\rangle\}$ and singlets. The latter are annihilated by both supercharges $Q$ and $Q^\dagger$ and have energy eigenvalue $E=0$.

The supercharges of the Nicolai model act in the Hilbert space of spin-less fermions on a 1D lattice. On a general lattice (or graph) $\Lambda$, spin-less fermions are described by creation and annihilation operators $c_i$, $c^\dagger_j$, satisfying
\[
\{ c_i, c^\dagger_j \} = \delta_{ij}, \qquad i,j \in \Lambda.
\]
The supercharge of the Nicolai model is given in eq. (\ref{eq:Qnic}) below.

\subsection{Supersymmetric lattice models}
In later years, many other supersymmetric lattice models for spin-less fermions have been proposed and studied.

\begin{description}
\item[M$_k$ models - integrable CFT and QFT \cite{FSdB, FNS}]
In these models, defined on 1D lattices, it is assumed that at the most $k$ fermions can occupy adjacent lattice sites. It has been found that these models can be tuned to be critical, and that their critical behavior is captured by the $k$-th model of ${\mathcal N}=2$ supersymmetric conformal field theory (CFT). Particular off-critical deformations, obtained by staggering parameters in the supercharges, lead to integrable ${\mathcal N}=2$ supersymmetric massive quantum field theories (QFT) with superspace superpotential given by Chebyshev polynomials \cite{FoS}.

\item[M$_1$ model on ladders and 2D lattices - superfrustration]
The M$_1$ model can be defined on any graph $\Lambda$. While for 1D chains the number of supersymmetric groundstates never exceeds 2, the number on 2D lattices tends to be exponential in the size (perimeter or area) of the lattice  \cite{vE,FeS05}.

\item[Models with $Q$ cubic in $c_i$]
These include the $\mathbb{Z}_2$ Nicolai model (with $g=0$) \cite{SKN17} and ${\mathcal N}=2$ supersymmetric SYK models \cite{FGMS}.
  
\item[Supersymmetric model of coupled fermion chains (FS model) \cite{FeS06}]
In these models the supercharge Q transports a particle from one chain to the other. To guarantee the fermionic nature of the supercharge, the particles on the individual chains are viewed as semions with fermion number $\pm 1/2$.

\item[A particle-hole symmetric version of the 1D M$_1$ model \cite{GFNR}]
Surprisingly, this model turns out to be equivalent to the FS model - an explicit unitary map was presented in \cite{FGGS}. 
\end{description}

A common feature to many of these models are large degeneracies of $E=0$ supersymmetric groundstates. In many cases it has been established that their number is exponential in the size of the system, meaning that the ground state entropy is extensive. It has been suggested \cite{Mo,PRTT} that such situations lead to a breaking of ergodicity and to a phenomenology similar to that of many-body localization (MBL).

\subsection{Counting supersymmetric groundstates}

Clearly, an important step in the analysis of supersymmetric lattice models is to understand the number and the nature of their supersymmetric ground states. This problem turns out to be quite hard in general. 

For the M$_k$ models in 1D a detailed understanding has been reached, thanks to integrability by Bethe Ansatz and to connections with supersymmetric CFT and (integrable) QFT \cite{FNS,FoS}. In the coupled chain model \cite{FeS06} the supersymmetric ground states are understood as tightly bound interchain pairs. In the equivalent particle-hole symmetric M$_1$ model the ground states have been analyzed with the help of a Bethe Ansatz \cite{GFNR}.

The ground state counting problem for the M$_1$ model on 2D lattices is understood in special cases \cite{vE,FeS05,HHFS,HS} but remains an open problem in general \cite{HMMSV}. An important observation \cite{FeS05} is that, from a mathematical perspective, the number of supersymmetric ground states is the dimension of the homology of $Q$. With this, the counting problem can be cast in strict mathematical terms, and mathematical tools for computing homologies can be employed.

For the Nicolai model, numerical analysis revealed ground state degeneracies for small system size, but the systematics behind these numbers remained unclear. The paper \cite{KMN} zoomed in a subset of all supersymmetric ground states, the so-called classical ground states. For the $\mathbb{Z}_2$ Nicolai model a recursion for the ground state generating function was presented in \cite{SKN17}.  In this letter we prove this conjectured relation and similarly establish a recursion for the generating function of the original Nicolai model. The working horse of our analysis is the homological perturbation lemma, which we present in section \ref{sec:homperlem}. 
Note that in order to rigourously present the derivation of the recursion relations for the ground state generating functions we have to switch to a quite formal language in the rest of the text.

\section{Homological computations} In this section we formulate the problem of the computation of the ground state generating functions for the one dimensional Nicolai and $\mathbb{Z}_2$ Nicolai models in formal purely mathematical terms and derive recursions for these functions.

By $\mathcal{H}_{n}$, $n\geq 0$, we denote the free graded commutative associative algebra generated by the degree $1$ elements $c_i^\dagger$, $i=1,\dots,n$. Its dimension is $2^n$. The operator $c_i\colon \mathcal{H}_{n}\to \mathcal{H}_{n}$, $i=1,\dots,n$, acts as the derivative with respect to $c_i^\dagger$. 
The operators $c_i$ and the operators of multiplication by $c_i^\dagger$, $i=1,\dots,n$, are the same as the annihilation and the creation operators given in the introduction, respectively. Abusing notation, we denote by $c_i^\dagger$ also the operator of multiplication by $c_i^\dagger$.

\subsection{Nicolai model}
The state space of the Nicolai model is $\mathcal{H}_{2m+1}$, $m\geq 1$. 
We consider the operator $Q\colon \mathcal{H}_{2m+1}\to \mathcal{H}_{2m+1}$ of degree $-1$ defined as
\[
Q:=\sum_{i=1}^m c_{2i-1} c_{2i}^\dagger c_{2i+1}.
\label{eq:Qnic}
\]
Obviously, $Q^2=0$, so we can compute its homology.  The ground state generating function of this model, $P_{2m+1}(z)$ is the Poincar\'e polynomial of the homology of $Q$, that is, 
\[
P_{2m+1}(z):=\sum_{i=1}^{2m+1} \dim H_i(\mathcal{H}_{2m+1},Q) z^i.
\] 

\begin{theorem} \label{thm:Nic} The polynomials $P_{2m+1}(z)$, $m\geq 3$, can be determined by the recursion
	\begin{align}\label{eq:nicolaigenfun}
	P_{2m+1}(z) = (1+z^2)P_{2m-1}(z) + (z + 2z^2+z^3)P_{2m-3}(z)
	\end{align}
	with the initial values given by
	\begin{align}\label{eq:nicolaiinitial}
	P_3(z) = 1 + 2z + 2z^2 + z^3
	\quad\text{and}\quad
	P_5(z) = 1 + 3z + 6z^2 + 6z^3 + 3z^4 + z^5.
	\end{align}
\end{theorem}

\begin{corollary} The total number of the ground states, $a_{2m+1}:= P_{2m+1}(1)$, satisfies the recursion 
	\[
	a_{2m+1} = 2 a_{2m-1} + 4 a_{2m-3}
	\]
	with the initial values given by $a_3=6$ and $a_5=20$. 
\end{corollary}

\subsection{$\mathbb{Z}_2$ Nicolai model}
The state space of the $\mathbb{Z}_2$ Nicolai model is $\mathcal{H}_{n}$, $n\geq 3$. We consider the operator $Q^{\mathbb{Z}_2}\colon \mathcal{H}_{n}\to \mathcal{H}_{n}$ of degree $-3$ defined as
\[
Q^{\mathbb{Z}_2}:=\sum_{i=1}^{n-2} c_{i} c_{i+1} c_{i+2}.
\]
Though $Q^{\mathbb{Z}_2}$ is not of degree $-1$, we still have $(Q^{\mathbb{Z}_2})^2=0$, so we can compute its homology.  The ground state generating function of this model, $P^{\mathbb{Z}_2}_{n}(z)$ is the Poincar\'e polynomial of the homology of $Q^{\mathbb{Z}_2}$, that is, 
\[
P^{\mathbb{Z}_2}_{n}(z):=\sum_{i=1}^{2m+1} \dim H_i(\mathcal{H}_n,Q^{\mathbb{Z}_2}) z^i.
\] 

\begin{theorem} \label{thm:Z2Nic} The polynomials $P^{\mathbb{Z}_2}_{n}(z)$, $n\geq 3$, can be determined by the recursion
	\begin{align}\label{eq:z2nicolaigenfun}
	P^{\mathbb{Z}_2}_{n}(z) = 2zP_{n-2}(z) + (z + z^2)P^{\mathbb{Z}_2}_{n-3}(z)
	\end{align}
	with the initial values given by
	\begin{align}\label{eq:z2nicolaiinitial}
	P^{\mathbb{Z}_2}_0(z) := 1,
	\qquad
	P^{\mathbb{Z}_2}_1(z) := 1 + z
	\quad\text{and}\quad
	P^{\mathbb{Z}_2}_2(z) = 1 + 2z + z^2.
	\end{align}
\end{theorem}

\begin{corollary}[Conjecture of Sannomyia, Katsura, and Nakayama \cite{SKN17}] The total number of the ground states, $a^{\mathbb{Z}_2}_{n}:= P^{\mathbb{Z}_2}_{n}(1)$, satisfies the recursion 
	\[
	a^{\mathbb{Z}_2}_{n} = 2 a^{\mathbb{Z}_2}_{n-2} + 2 a^{\mathbb{Z}_2}_{n-3}
	\]
	with the initial values given by $a^{\mathbb{Z}_2}_0=1$, $a^{\mathbb{Z}_2}_1=2$, and $a^{\mathbb{Z}_2}_2=4$. 
\end{corollary}

\subsection{Homological perturbation lemma} \label{sec:homperlem} The main technical tool that we use in the proofs of Theorems~\ref{thm:Nic} and~\ref{thm:Z2Nic} is a version 
of the so-called homological perturbation lemma. Consider a graded vector space $\mathcal{H}$ with two commuting differentials of degree $-1$, $d_1$ and $d_2$. Assume that we have chosen a deformation retract data connecting the differential graded spaces $\mathcal{H}$ with the differential $d_1$ and its homology, that is, the space $H_{\bullet}(\mathcal{H},d_1)$ with the zero differential. By a deformation retract data we mean that we have an operator $h\colon \mathcal{H}\to \mathcal{H}$ and two quasi-isomorphism (the chain maps that induce the isomorphisms on the homology level) $i\colon H_{\bullet}(\mathcal{H},d_1) \to \mathcal{H}$ and $p\colon \mathcal{H}\to H_{\bullet}(\mathcal{H},d_1)$, which can be arranged in a diagram
\begin{center}
	\begin{tikzpicture}
	\path node (A) {$(H_{\bullet}(\mathcal{H},d_1),0)$}
	(0:3cm)    node (B) {$(\mathcal{H},d_1)$};
	\path[-stealth]
	([yshift= 2.5pt]B.west) edge node [below,yshift=-1.0ex]  {$i$} ([yshift= 2.5pt]A.east)
	([yshift= -2.5pt]A.east) edge node [above,yshift= 1.0ex]  {$p$} ([yshift= -2.5pt]B.west);
	\path[-stealth] (B) edge  [out=10,in=-10,looseness=8] node[right] {$h$} (B);
	\end{tikzpicture}
\end{center}
 such that
\begin{align}
pi = \mathrm{Id}_{H_{\bullet}(\mathcal{H},d_1)} \qquad \text{and} \qquad ip = \mathrm{Id}_{\mathcal{H}} + d_1h+ hd_1
\end{align}
(note that the condition $pi = \mathrm{Id}_{H_{\bullet}(\mathcal{H},d_1)}$ just follows from the assumption that both $p$ and $i$ are quasi-isomorphisms, so we just included it here for completeness), and the operator $1-d_2h$ is invertible (for instance, there can be an additional gradation that guarantees invertibility, as in the case of bicomplex). 

\begin{lemma}[\cite{Brown}]\label{lem:HomologicalPerturbation}
	Under these assumptions we have the following isomorphism of graded vector spaces: 
	\[
	H_{\bullet}(\mathcal{H},d_1+d_2) \cong H_\bullet(H_{\bullet}(\mathcal{H},d_1), p(1-d_2h)^{-1}d_2i). 
	\]
\end{lemma}

This lemma has many much stronger versions and refinements, but this form is exactly what we use in this paper. 

One more definition that will be useful below is the suspension of a chain complex.

\begin{definition}\label{def:suspension}
	Let $(\mathcal{H},Q)$ be a chain complex. For any $k\in\mathbb{Z}$ the chain complex $(\mathcal{H}[k],Q^{[k]})$ is defined by formally adding $k$ to the gradation,
	and the differential $Q$ is twisted by the sign $(-1)^k$. On the level of Poincar\'e polynomials the effect of suspension corresponds to multiplication by $z^k$.
\end{definition}

\subsection{Proof of Theorem~\ref{thm:Nic}} We define the differentials $d_1:=c_{2m-1}c_{2m}^\dagger c_{2m+1}$ and $d_2:=Q-d_1$. We have:
\[
H_{\bullet} (\mathcal{H}_{2m+1},d_1) \cong
\langle 1, c_{2m-1}^\dagger, c_{2m+1}^\dagger,
c_{2m-1}^\dagger c_{2m}^\dagger,
c_{2m}^\dagger c_{2m+1}^\dagger,
c_{2m-1}^\dagger c_{2m}^\dagger c_{2m+1}^\dagger
\rangle\cdot\mathcal{H}_{2m-2}
\]
The choice of the representatives of the homology classes of $d_1$ here defines a natural quasi-isomorphism 
$i\colon H_{\bullet} (\mathcal{H}_{2m+1},d_1) \to \mathcal{H}_{2m+1}$.
Define the map $h\colon  \mathcal{H}_{2m+1} \to \mathcal{H}_{2m+1}$ as $h:=c_{2m+1}^\dagger c_{2m} c_{2m-1}^\dagger$, and the map $$
p\colon \mathcal{H}_{2m+1} \to \langle 1, c_{2m-1}^\dagger,  c_{2m+1}^\dagger,
c_{2m-1}^\dagger c_{2m}^\dagger,
c_{2m}^\dagger c_{2m+1}^\dagger,
c_{2m-1}^\dagger c_{2m}^\dagger c_{2m+1}^\dagger
\rangle\cdot\mathcal{H}_{2m-2}
\subset \mathcal{H}_{2m+1}
$$ 
as the projection with the kernel $\langle c_{2m}^\dagger, 
c_{2m-1}^\dagger c_{2m+1}^\dagger,
\rangle\cdot\mathcal{H}_{2m-2}$.

The maps $h$, $i$, and $p$ satisfy all conditions of Lemma~\ref{lem:HomologicalPerturbation}.
In particular, the inverse of $1-d_2h$ is $1+d_2h$ (it is straightforward to check that $hd_2h=0$). Thus we have:
\[
H_{\bullet} (\mathcal{H}_{2m+1},Q)\cong
H_{\bullet} (\langle 1, c_{2m-1}^\dagger, c_{2m+1}^\dagger,
c_{2m-1}^\dagger c_{2m}^\dagger,
c_{2m}^\dagger c_{2m+1}^\dagger,
c_{2m-1}^\dagger c_{2m}^\dagger c_{2m+1}^\dagger
\rangle\cdot\mathcal{H}_{2m-2},
p d_2 i + p d_2 h d_2 i )
\]
(note that the operator $i$ here acts tautologically, but we keep it for the sake of notation). 

Observe that the only summand in $d_2$ that can affect the generators $c_{2m-1}^\dagger$, $c_{2m}^\dagger$, and $c_{2m+1}^\dagger$ in $\mathcal{H}_{2m+1}$ is $c_{2m-3} c_{2m-2}^\dagger c_{2m-1}$. The image of $c_{2m-3} c_{2m-2}^\dagger c_{2m-1}i$ is a subspace of 
\[
\langle c_{2m-2}^\dagger, c_{2m-2}^\dagger c_{2m}^\dagger , c_{2m-2}^\dagger c_{2m}^\dagger c_{2m+1}^\dagger \rangle \cdot \mathcal{H}_{2m-4},
\]
 therefore, the images of $hd_2i$ and $d_2hd_2i$ are subspaces of $c_{2m+1}^\dagger c_{2m-1}^\dagger c_{2m-2}^\dagger \mathcal{H}_{2m-4}$, which lies in the kernel of the projection $p$. So, the term $ p d_2 h d_2 i $ acts trivially on the homology of $d_1$. Therefore,
\[
H_{\bullet} (\mathcal{H}_{2m+1},Q)\cong
H_{\bullet} (\langle 1, c_{2m-1}^\dagger, c_{2m+1}^\dagger,
c_{2m-1}^\dagger c_{2m}^\dagger,
c_{2m}^\dagger c_{2m+1}^\dagger,
c_{2m-1}^\dagger c_{2m}^\dagger c_{2m+1}^\dagger
\rangle\cdot\mathcal{H}_{2m-2},
p d_2 i )
\]

We split the space $\langle 1, c_{2m-1}^\dagger, c_{2m+1}^\dagger,
c_{2m-1}^\dagger c_{2m}^\dagger,
c_{2m}^\dagger c_{2m+1}^\dagger,
c_{2m-1}^\dagger c_{2m}^\dagger c_{2m+1}^\dagger
\rangle\cdot\mathcal{H}_{2m-2}$ into the eigenspaces of the action of $pd_2i$ as 
\[
\langle 1, c_{2m-1}^\dagger \rangle\cdot\mathcal{H}_{2m-2}
\oplus 
\langle 
c_{2m}^\dagger c_{2m+1}^\dagger,
c_{2m-1}^\dagger c_{2m}^\dagger c_{2m+1}^\dagger
\rangle\cdot\mathcal{H}_{2m-2}
\oplus 
\langle c_{2m+1}^\dagger\rangle\cdot\mathcal{H}_{2m-2}
\oplus
\langle c_{2m-1}^\dagger c_{2m}^\dagger\rangle\cdot\mathcal{H}_{2m-2}
\]
From the definition of $d_2$ we have:
\begin{align}
(\langle 1, c_{2m-1}^\dagger \rangle\cdot\mathcal{H}_{2m-2}, pd_2i) & \cong (\mathcal{H}_{2m-1},Q )
\\
(\langle 
c_{2m}^\dagger c_{2m+1}^\dagger,
c_{2m-1}^\dagger c_{2m}^\dagger c_{2m+1}^\dagger
\rangle\cdot\mathcal{H}_{2m-2}, pd_2i) & \cong 
(\mathcal{H}_{2m-1}[2],Q^{[2]}).
\end{align}
As for the last two summands, we note that $pc_{2m-3} c_{2m-2}^\dagger c_{2m-1}i$ acts trivially on $\langle c_{2m+1}^\dagger\rangle\cdot\mathcal{H}_{2m-2}$ and $\langle c_{2m-1}^\dagger c_{2m}^\dagger\rangle\cdot\mathcal{H}_{2m-2}$, therefore 
\begin{align}
(\langle c_{2m+1}^\dagger\rangle\cdot\mathcal{H}_{2m-2}, pd_2i) & \cong
\langle c_{2m+1}^\dagger, c_{2m-2}^\dagger c_{2m+1}^\dagger\rangle\cdot\mathcal{H}_{2m-3}, pd_2i) 
\\
& \cong (\mathcal{H}_{2m-3}[1],Q^{[1]} )\oplus (\mathcal{H}_{2m-3}[2],Q^{[2]} )
\\
(\langle c_{2m-1}^\dagger c_{2m}^\dagger\rangle\cdot\mathcal{H}_{2m-2}, pd_2i) 
& \cong (\langle c_{2m-1}^\dagger c_{2m}^\dagger, c_{2m-2}^\dagger c_{2m-1}^\dagger c_{2m}^\dagger \rangle\cdot\mathcal{H}_{2m-3}, pd_2i) 
\\
& \cong (\mathcal{H}_{2m-3}[2],Q^{[2]} )\oplus (\mathcal{H}_{2m-3}[3],Q^{[3]} )
\end{align}

Thus we prove that $H_{\bullet} (\mathcal{H}_{2m+1},Q)$ is isomorphic to the homology of the direct sum of complexes 
\[
(\mathcal{H}_{2m-1},Q ) \oplus (\mathcal{H}_{2m-1}[2],Q^{[2]})
\oplus (\mathcal{H}_{2m-3}[1],Q^{[1]} )\oplus (\mathcal{H}_{2m-3}[2],Q^{[2]} )
\oplus  (\mathcal{H}_{2m-3}[2],Q^{[2]} )\oplus (\mathcal{H}_{2m-3}[3],Q^{[3]} ).
\]
Therefore, the Poincar\'e polynomial $P_{2m+1}(z)$ of the homology  of $(\mathcal{H}_{2m+1},Q)$ is equal to
\[
P_{2m-1}(z) + z^2P_{2m-1}(z)
+ zP_{2m-3}(z) + z^2P_{2m-3}(z) + z^2 P_{2m-3}(z) + z^3 P_{2m-3}(z).
\]
This completes the proof of the Theorem~\ref{thm:Nic}.

\subsection{Proof of Theorem~\ref{thm:Z2Nic}} In this case the operator $Q^{\mathbb{Z}_2}$ is not of degree $-1$, so we have two ways how to turn this situation into a standard one: either we can just redevelop the homological perturbation lemma for the operators of degree $-3$ (then the homotopy contraction operator $h$ should be of degree $+3$), of we can split the space $\mathcal{H}_n$ into the direct sum of three subspaces, depending on the remainder $\!\mod 3$ of the degree, and redefine the degree for each of them as the quotient of division with remainder by $3$ of the original degree. These two approaches are equivalent, and we choose the first one since the ground state generating function is formulated in terms of the original degrees.  

We define the differentials $d_1:=c_{n-2}c_{n-1}c_n$ and $d_2:=Q^{\mathbb{Z}_2}-d_1$. We have:
\[
H_\bullet(\mathcal{H}_n,d_1)\cong \langle c_{n-2}^\dagger, c_{n-1}^\dagger, c_{n}^\dagger, c_{n-2}^\dagger c_{n-1}^\dagger, c_{n-2}^\dagger c_{n}^\dagger, c_{n-1}^\dagger c_{n}^\dagger \rangle \cdot \mathcal{H}_{n-3},
\]
and this isomorphism defines a natural map $i\colon H_\bullet(\mathcal{H}_n,d_1) \to \mathcal{H}_n$. Define the map $h\colon \mathcal{H}_n\to \mathcal{H}_n$ as $h:=c_n^\dagger c_{n-1}^\dagger c_{n-2}^\dagger$, and the map
\[
p\colon \mathcal{H}_n \to \langle c_{n-2}^\dagger, c_{n-1}^\dagger, c_{n}^\dagger, c_{n-2}^\dagger c_{n-1}^\dagger, c_{n-2}^\dagger c_{n}^\dagger, c_{n-1}^\dagger c_{n}^\dagger \rangle \cdot \mathcal{H}_{n-3}\subset \mathcal{H}_n
\]
as the projection with kernel $\langle 1, c_n^\dagger c_{n-1}^\dagger c_{n-2}^\dagger \rangle \cdot \mathcal{H}_{n-3}$. 

The maps $h$, $i$, and $p$  satisfy all conditions of Lemma~\ref{lem:HomologicalPerturbation}, and since $hd_2h=0$, the inverse of $1-d_2h$ is equal to $1+d_2h$. Thus we have:
\[
H_{\bullet} (\mathcal{H}_{n},Q^{\mathbb{Z}_2})\cong
H_{\bullet} (\langle c_{n-2}^\dagger, c_{n-1}^\dagger, c_{n}^\dagger, c_{n-2}^\dagger c_{n-1}^\dagger, c_{n-2}^\dagger c_{n}^\dagger, c_{n-1}^\dagger c_{n}^\dagger \rangle \cdot \mathcal{H}_{n-3},
p d_2 i + p d_2 h d_2 i )
\]
(note that the operator $i$ here act tautologically, but we keep it for the sake of notation). 

Observe that the only  two summands in $d_2$ that can affect the generators $c_{n-2}^\dagger$, $c_{n-1}^\dagger$, and $c_{n}^\dagger$ in $\mathcal{H}_{n}$ are 
$c_{n-4}c_{n-3}c_{n-2}+c_{n-3}c_{n-2}c_{n-1}$. The image of $(c_{n-4}c_{n-3}c_{n-2}+c_{n-3}c_{n-2}c_{n-1})i$ is a subspace of 
\[
\mathcal{H}_{n-4} \oplus \langle c_{n-1}^\dagger, c_n^\dagger\rangle \cdot \mathcal{H}_{n-5},
\]
so the images of $hd_2i$ and $d_2hd_2i$ are subspaces of $c_{n-2}^\dagger c_{n-1}^\dagger c_n^\dagger \cdot \mathcal{H}_{n-4}$, which lies in the kernel of the projection $p$. So, the term $ p d_2 h d_2 i $ acts trivially on the homology of $d_1$. Therefore, 
\[
H_{\bullet} (\mathcal{H}_{n},Q^{\mathbb{Z}_2})\cong
H_{\bullet} (\langle c_{n-2}^\dagger, c_{n-1}^\dagger, c_{n}^\dagger, c_{n-2}^\dagger c_{n-1}^\dagger, c_{n-2}^\dagger c_{n}^\dagger, c_{n-1}^\dagger c_{n}^\dagger \rangle \cdot \mathcal{H}_{n-3}, p d_2 i ).
\]
We split the space
$\langle c_{n-2}^\dagger, c_{n-1}^\dagger, c_{n}^\dagger, c_{n-2}^\dagger c_{n-1}^\dagger, c_{n-2}^\dagger c_{n}^\dagger, c_{n-1}^\dagger c_{n}^\dagger \rangle \cdot \mathcal{H}_{n-3}$
into a direct sum of subspaces invariant under the action of $pd_2i$ as
\[
\langle c_{n}^\dagger,  c_{n-2}^\dagger c_{n}^\dagger \rangle \cdot \mathcal{H}_{n-3}
\oplus
\langle c_{n-1}^\dagger,  c_{n-2}^\dagger c_{n-1}^\dagger \rangle \cdot \mathcal{H}_{n-3}
\oplus
\langle c_{n-2}^\dagger\rangle \cdot \mathcal{H}_{n-3}
\oplus
\langle c_{n-1}^\dagger c_{n}^\dagger\rangle \cdot \mathcal{H}_{n-3}.
\]
On the first two spaces the operator $pc_{n-3}c_{n-2}c_{n-1}i$ acts trivially. Then, since 
$\langle c_{n}^\dagger,  c_{n-2}^\dagger c_{n}^\dagger \rangle \cdot \mathcal{H}_{n-3}\cong \langle c_{n}^\dagger \rangle \cdot \mathcal{H}_{n-2}$ and
$\langle c_{n-1}^\dagger,  c_{n-2}^\dagger c_{n-1}^\dagger \rangle \cdot \mathcal{H}_{n-3} \cong \langle c_{n-1}^\dagger \rangle \cdot \mathcal{H}_{n-2}$, we have 
\begin{align}
(\langle c_{n}^\dagger,  c_{n-2}^\dagger c_{n}^\dagger \rangle \cdot \mathcal{H}_{n-3}, pd_2i)
& \cong (\mathcal{H}_{n-2}[1], (Q^{\mathbb{Z}_2})^{[1]}), \\
(\langle c_{n-1}^\dagger,  c_{n-2}^\dagger c_{n-1}^\dagger \rangle \cdot \mathcal{H}_{n-3}, pd_2i)
& \cong (\mathcal{H}_{n-2}[1], (Q^{\mathbb{Z}_2})^{[1]}).
\end{align}
On $\langle c_{n-2}^\dagger\rangle \cdot \mathcal{H}_{n-3}$ and 
$\langle c_{n-1}^\dagger c_{n}^\dagger\rangle \cdot \mathcal{H}_{n-3}$ the operator
$p(c_{n-4}c_{n-3}c_{n-2}+c_{n-3}c_{n-2}c_{n-1})i$ acts trivially, therefore,
\begin{align}
(\langle c_{n-2}^\dagger\rangle \cdot \mathcal{H}_{n-3}, pd_2i)
& \cong (\mathcal{H}_{n-3}[1], (Q^{\mathbb{Z}_2})^{[1]}), \\
(
\langle c_{n-1}^\dagger c_{n}^\dagger\rangle \cdot \mathcal{H}_{n-3}, pd_2i)
& \cong (\mathcal{H}_{n-3}[2], (Q^{\mathbb{Z}_2})^{[2]}).
\end{align}
Thus we prove that $H_{\bullet} (\mathcal{H}_{n},Q^{\mathbb{Z}_2})$ is isomorphic to the homology of 
\[
(\mathcal{H}_{n-2}[1], (Q^{\mathbb{Z}_2})^{[1]})\oplus  (\mathcal{H}_{n-2}[1], (Q^{\mathbb{Z}_2})^{[1]})  \oplus (\mathcal{H}_{n-3}[1], (Q^{\mathbb{Z}_2})^{[1]}) \oplus
(\mathcal{H}_{n-3}[2], (Q^{\mathbb{Z}_2})^{[2]}) .
\]
Therefore, the Poincar\'e polynomial $P_n^{\mathbb{Z}_2}(z)$ of the homology of $(\mathcal{H}_{n},Q^{\mathbb{Z}_2})$ is equal to
\[
zP_{n-2}^{\mathbb{Z}_2}(z)+zP_{n-2}^{\mathbb{Z}_2}(z)+zP_{n-3}^{\mathbb{Z}_2}(z)+z^2P_{n-2}^{\mathbb{Z}_2}(z).
\]
This completes the proof of the Theorem~\ref{thm:Z2Nic}.

\section{Acknowledgements}
We thank H.~Katsura for sharing his insights on this subject. S.~S. was supported by the Netherlands Organization for Scientific Research.

\end{document}